\def\be{\begin{equation}}
\def\ee{\end{equation}}
\def\bea{\begin{eqnarray}}
\def\eea{\end{eqnarray}}

\documentclass[12pt]{article}
\usepackage{epsfig,latexsym}

\begin{document}
\thispagestyle{empty}
\vspace*{-0.5 cm}
\vspace*{-1.2in}
\begin{flushright}
{INFCT 01/02} \\ Feb-2002 \\
\end{flushright}
\vspace*{0.7 in}
\begin{center}
{\large \bf Perturbative and non-perturbative aspects of the proper time renormalization group}
\\
\vspace*{1cm} 
{\bf D. Zappal\`a }\\ \vspace*{.3cm} 
{\it INFN, Sezione di Catania}\\
{\it Dipartimento di Fisica, Universit\`a di Catania}\\
{\it Corso Italia 57, I-95129, Catania, Italy} \\

\vspace*{1 cm}
{\bf ABSTRACT} \\
\end{center}
The renormalization group flow equation obtained by means of a 
proper time regulator is used to calculate the two loop 
$\beta$-function and anomalous dimension $\eta$ of the field for the 
$O(N)$ symmetric scalar theory.  The standard perturbative analysis 
of the flow equation does not yield the correct results for both  
$\beta$ and $\eta$.  We also show that it is still possible to extract 
the correct $\beta$ and $\eta$ from the flow equation in a 
particular limit of the infrared scale.
A modification of the derivation of the Exact Renormalization 
Group flow,  which involves  a more general class of regulators, 
to recover the  proper time renormalization group flow is analyzed.
\\
\vskip 0.5 cm
\noindent
Pacs 11.10.Gh , 11.10.Hi

\parskip 0.3 cm
\vspace*{3cm}
\setcounter{page}{1}
\voffset -1in
\vskip2.0cm

\section{Introduction}

The description of critical phenomena and, in general, of phenomena which involve the physics of 
many scales at the same time,  is successfully achieved  by means of the Renormalization Group 
techniques developed on the basis of the work of Wilson\cite{kad}. Recently these techniques have been 
refined by many authors and a particular version of the flow equation, generally indicated
as Exact Renormalization Group (ERG) has been formulated 
\cite{polch,wet1,wetplb,bdm,ellw,mor1}. 
Basically this flow is called exact because it is formally derived from the generator of 
the Green functions of the theory, without explicitly performing the functional integration and 
without resorting to any specific approximation or truncation. Many details and applications 
of this flow equation  can be found for instance in \cite{berg,bagn}.

Beside the ERG, other  formulations of the renormalization group flow have been derived.
In particular  we shall focus on one of them which is known as Proper Time Renormalization
Group (PTRG) and which has been formulated and studied in many applications 
\cite{ole,flo,sen1,sha1,boza,maza,za,liti1,liti2,lpmain} .
This flow equation is obtained by making use of the Proper Time representation\cite{schw} 
and by introducing a suitable multiplicative regulator as a cut-off on 
the Infrared (IR)  modes. It must be remarked that, substantially, the PTRG is obtained in the 
quoted papers as a  renormalization group  improvement of the  one loop effective action. 

Due to the lack of a formal proof of exactness similar to the one concerning the ERG and 
due to its straightforward interpretation in terms of improvement on the one loop calculation,
many doubts have been cast about the complete reliability of the results obtained from the 
PTRG equation. On the other hand there are some examples where this equation  provided 
reliable results. 
In fact the derivative expansion truncated to the lowest order (local potential approximation)
and to the first order (which consists of two coupled flow equations for the potential $U_k$
and for the wave function renormalization $Z_k$) has been employed to determine some critical
exponents at the non-gaussian fixed point of the scalar theory in three dimensions\cite{boza,maza}.
These coupled flow equations  have been used for another application in quantum mechanics,
namely the determination of the energy gap between the ground state and the first excited state 
of the double well potential\cite{za}, where the exact value can be numerically evaluated and 
therefore a quantitative check of the flow equations is possible.  Obviously these results 
determined by means of the PTRG show a dependence on the form of the proper time regulator 
considered and it turned out that the optimal choice corresponds to the  sharp limit of  this
regulator.  In this limit the flow equations for $U_k$ and $Z_k$ have a simple form and 
the various quantities determined are certainly comparable to the ones extracted from the ERG.

Then it would be interesting to analyze the relation between this flow and the ERG. This 
issue has already been addressed in \cite{liti1,liti2,lpmain}.  In this paper we focus both on
some properties of the perturbative expansion of the PTRG and on the possibility of 
introducing a cut-off on the ultraviolet modes for the PTRG, with the same
procedure employed for  the ERG.
Before illustrating the details of our analysis we examine some points concerning the 
Proper Time flow equation and the  sharp limit on the regulator which has 
been mentioned above. The Proper Time flow equation  used  in \cite{za} is
\be
k\partial_k\Gamma_k=-\frac{1}{2}\mathrm{Tr}\int_0^\infty\;\;\;\frac{\mathrm{d}s}{s}\,
(k\partial_kf_k)\exp{\left(-s\frac{\delta^2 \Gamma_k}{\delta\varphi\delta\varphi}\right)}
\ee
where the derivative of the regulator $k\partial_kf_k$ has the form
\be\label
{eq:dercut}
k \partial_kf_k=-2smZ_kk^2\,\exp(-smZ_kk^2)\frac{(smZ_kk^2)^m}{m!}
\ee
and $Z_k$ is the wave function renormalization coefficient of the kinetic term 
$(\partial \varphi)^2$ of the running effective action $\Gamma_k$ and $m$ is a positive integer.
The regulator considered here depends on the particular value assigned to the index $m$
and, as discussed before it has been checked that in the various applications the results 
are optimized in the sharp limit which corresponds to $m\to \infty$ \cite{boza}
(incidentally we notice that a slightly different form of $k \partial_kf_k$ has been used in 
\cite{boza,maza} and by other authors, which however is less suitable to study the 
limit $m\to \infty$). In \cite{za} the flow equation for $U_k$ and $Z_k$ were 
obtained in this limit but  here we are interested in the form of the full flow equation.
To analyze the large $m$ limit we follow the derivation outlined in \cite{liti2}. We 
replace the factorial in Eq. (\ref{eq:dercut}) by making
use of the Stirling formula 
$(m-1) !=e^{-m} m^m\sqrt{(2\pi/m) }(1+O(1/m))$
and neglecting the terms that  vanish in the limit $m \to\infty$ we get
\be\label{eq:poststir}
k \partial_k f_k=
-2sZ_kk^2\,\exp\left[-m(sZ_kk^2-\ln (sZ_kk^2)-1)\right]\sqrt\frac{m}{2\pi}
\ee
The limit  $m\to\infty$ then yields 
\bea
\label{limite}
&&
\lim_{m\to\infty}k\partial_kf_k^m=
-\sqrt{2}sZ_kk^2\,\delta\left(\sqrt{sZ_kk^2-\ln (sZ_kk^2)-1}\right)
\nonumber \\
&&
=-2 s \; \delta\left(s-\frac{1}{Z_k k^2}\right)
\eea
where, in the first step we have introduced the $\delta$-function according to 
\be 
\label{delt}
\lim_{\epsilon\to 0} \frac{ {\rm exp}\left (-x^2/ \epsilon^2 \right )}{\epsilon \sqrt{\pi}}=\delta(x)
\ee
and with the position $m=1/\epsilon^2$.
The result in Eq. (\ref{limite}) directly leads to the full flow equation in the 
large $m$ limit which,  according to  Eq. (\ref{limite}), reads
\be
\label{eq:gamma}
k\;{\partial \Gamma_k \over \partial k }=
{\rm Tr} \; \left \{ 
{\rm exp}\left  (-\frac{1}{Z_k k^2}\;\;  {\delta^2 \Gamma_k \over 
\delta \varphi\delta \varphi} \right ) \right\}  
\ee
Eq. (\ref{eq:gamma}) is the starting point of our analysis and 
in the following we indicate this flow equation as the PTRG flow.
Clearly the derivative expansion of Eq. (\ref{eq:gamma}) yields again 
those flow equation  for $U_k$ and $Z_k$ which have been analyzed 
in \cite{maza,za}.

In Section 2 we consider a purely perturbative calculation 
and we evaluate by means of the flow equation the two loop anomalous dimension 
of the field and $\beta$-function for the $O(N)$ symmetric scalar theory in 
four dimensions, which, as it is well known, correspond respectively to the 
$O(\lambda^2)$ and  $O(\lambda^3)$  terms in the expansion in powers of 
the quartic coupling $\lambda$. This calculation has already been performed 
by making use of the ERG \cite{pw},
but it is interesting to reconsider it within the PTRG framework because of
the recent  analysis in \cite{lpmain} where it is argued that the proper time 
flow equation does not respect  the diagrammatic structure 
beyond one loop and should therefore produce wrong predictions 
within a perturbative expansion. We develop a systematic procedure
to calculate the one-particle irreducible (1PI) vertices in the perturbative 
expansion and show that the correct two loop  anomalous dimension and 
$\beta$-function are not reproduced by the standard 
perturbative analysis of Eq. (\ref{eq:gamma}). However we also discuss a
particular limit of the flow equation in which the correct two loop 
quantities are recovered. 
In Section 3 we follow the procedure outlined in \cite{wetplb} to derive 
the ERG flow and try to reformulate it in the case of the PTRG. 
Finally, Section 4 contains some comments about the results 
obtained and the conclusions.

\section{The two loop $\beta$-function}

In this Section we consider the perturbative analysis of Eq. (\ref{eq:gamma}) and, 
in particular, we discuss the computation of the anomalous dimension $\eta$  and the 
$\beta$-function of the $O(N)$ scalar theory respectively to  order $O(\lambda^2)$ and 
$O(\lambda^3)$  of their expansion in powers of the quartic coupling 
constant $\lambda$, which correspond to the  two loop calculation of these quantities.
We shall check that the usual correct perturbative procedure fails at two loop but at 
the same time we show that the two loop $\beta$ and $\eta$ are recovered from  
Eq. (\ref{eq:gamma}) in a specific  limit.

To this aim we need the coupled  flow  equations for various 1PI vertices which, 
in turn, must be obtained from the full equation for $\Gamma_k$ by functional derivation. 
Once the set of (infinite) coupled flow equations for the 1PI vertices is determined, 
it can be solved within  a particular truncation which is dictated by the specific
order in the perturbative series one is interested in.

The first step concerns the functional derivation of the full flow equation which involves an exponential term.
The latter must be properly expanded in order to avoid commutation problems when taking the derivatives
(in the following we can safely neglect the factor $1/Z_k=1+O(\lambda)$ 
which appears in the exponential in Eq. (\ref{eq:gamma}) because it does not give any contribution 
to the perturbative order we are interested in).
Therefore it is convenient to disentangle in  $ (\delta^2 \Gamma_k / 
\delta \phi\delta \phi)$ the field independent and the field dependent parts which we call respectively
$A$ and $B$:
\be
\label{split}
{\delta^2 \Gamma_k \over 
\delta \phi\delta \phi} =A+B
\ee
\noindent
and then the required expansion, which has been derived in \cite{schw}, is
\bea
\label{schwexp}
&&{\rm Tr} \;\left \{ {\rm exp}\left (-\frac{  A+B}{k^2} \right )\right\}=
{\rm Tr} \; \left \{ {\rm exp}\left (-\frac{ A}{k^2} \right ) \right\}  
-\frac{1}{k^2} \; {\rm Tr} \;\left \{  B \cdot  {\rm exp}\; \left (-\frac{A}{k^2}  \right )\right\} \nonumber \\
&&
+\frac{1}{2 k^4} \; {\rm Tr} \; \Biggl \lbrace \int_0^1 {\rm d} u \Biggl \lbrack B \cdot {\rm exp}\;
\left (-\frac{ A}{k^2}(1-u) \right)\cdot 
B\cdot  {\rm exp}\left (-\frac{ A}{k^2} u  \right ) \Biggr \rbrack \Biggr \rbrace \nonumber \\
&&
+...+ \frac{(-1)^{n+1}}{(n+1) \;k^{2n+2}} {\rm Tr} \; \Biggl \lbrace   \int_0^1 {\rm d}u_1 ... 
\int_0^1 {\rm d}u_n \Biggl\lbrack\; u_1^{n-1} u_2^{n-2}... u_{n-1}^1 u_n^0\nonumber \\
&&
B\cdot {\rm exp}\left (-\frac{A}{k^2}(1-u_1)  \right )\cdot 
B\cdot  {\rm exp} \left  (-\frac{ A}{k^2}u_1(1-u_2) \right ) \cdot 
B \nonumber \\
&&
\cdot {\rm exp} \left  (-\frac{A}{k^2}u_1u_2(1-u_3) \right )\cdot B\cdot ... \cdot
B\cdot  {\rm exp}\left (-\frac{A}{k^2}u_1u_2...u_{n-1}(1-u_n) \right) \nonumber \\
&& 
\cdot B\cdot {\rm exp}\left (-\frac{A}{k^2}u_1u_2...u_{n-1}u_n \right) 
\Biggr \rbrack\Biggr \rbrace +...
\eea

In Eq. (\ref{schwexp}) only  the field independent part $A$ appears in the exponentials and it is now simple to 
take the functional derivatives of this expression where the field dependence is all contained in $B$.
The $n$-point 1PI function is obtained by evaluating the functional derivative of order $n$ of 
Eq. (\ref{schwexp}) for vanishing fields. Moreover, due to the splitting in Eq. (\ref{split}), in a 
perturbative expansion one expects that $A$ contains the lowest order term which is not proportional to 
the coupling constant whereas the field dependent part $B$ is a $O(\lambda)$ term and therefore 
Eq.  (\ref{schwexp}) naturally provides an expansion in powers of the coupling constant.

For our purpose we shall consider  the case of the quartic massless $O(N)$ scalar theory in four 
dimensions 
with Euclidean action
\be
\label{clact}
S=\int {\rm d}^4 x \left (\frac{1}{2}\partial \phi_i \partial \phi_i +\frac{\lambda}{4!}(\phi_i\phi_i)^2 \right )
\ee
Due to the symmetry, the $n$-point 1PI functions with odd $n$ are zero 
to each order of  the perturbative expansion. Therefore from  Eq. (\ref{schwexp})
one gets the following flow equations
in the momentum space for the two, four and six-point functions

\be
\label{gamma2}
k\;{\partial \Gamma^{(2)}_{i,j}(q,-q) \over \partial k }=-\frac{1}{k^2} \; {\rm Tr_t} 
\; \left \{  \Gamma^{(4)}_{i,j,l_1,l_2}(q,-q,t,-t) \; D^{1}_{l_2,l_1}(t) 
\right\} 
\ee

\vskip 30 pt

\bea\label{gamma4}
&&k\;{\partial \Gamma^{(4)}_{i,j,r,s}(q,-q,p,-p) \over \partial k }=
\frac{1}{k^4} \; {\rm Tr_t} 
\; \Biggl \lbrace   \int_0^1 {\rm d}u_1
\nonumber \\
&&
\times\Biggl \lbrack 
\Gamma^{(4)}_{i,r,l_1,l_2}(q,p,t,-q-p-t) D^{1-u_1}_{l_2,l_3}(q+p+t)  
\Gamma^{(4)}_{j,s,l_3,l_4}(-q,-p,q+p+t,-t) D^{u_1}_{l_4,l_1}(t) 
\nonumber \\
&&
+\Gamma^{(4)}_{j,r,l_1,l_2}(-q,p,t,q-p-t) D^{1-u_1}_{l_2,l_3}(-q+p+t)  
\Gamma^{(4)}_{i,s,l_3,l_4}(q,-p,-q+p+t,-t) D^{u_1}_{l_4,l_1}(t) 
\nonumber \\
&&
+\Gamma^{(4)}_{i,j,l_1,l_2}(q,-q,t,-t)\; D^{1-u_1}_{l_2,l_3}(t) \;
\Gamma^{(4)}_{r,s,l_3,l_4}(p,-p,t,-t)\; D^{u_1}_{l_4,l_1}(t) 
\Biggr \rbrack 
\Biggr \rbrace \nonumber \\
&&
-\frac{1}{k^2} \; {\rm Tr_t} 
\; \left \{  \Gamma^{(6)}_{i,j,r,s,l_1,l_2}(q,-q,p,-p,t,-t)\; D^{1}_{l_2,l_1}(t) \right \}
\eea

\vskip 30 pt

\bea\label{gamma6}
&&k\;{\partial \Gamma^{(6)}_{i,j,r,s,w,z}(0,0,0,0,q,-q) \over \partial k }=-\frac{1}{3k^6} \; {\rm Tr_t} 
\; \Biggl \lbrace  \int_0^1 {\rm d}u_1\;u_1 \int_0^1 {\rm d}u_2 \nonumber \\
&&\times\Biggl \lbrack  3 {\cal P}_{(i,j,r,s)} 
\Biggl (
\Gamma^{(4)}_{w,z,l_1,l_2}(q,-q,t,-t)\;D^{1-u_1}_{l_2,l_3}(t)\;
\Gamma^{(4)}_{i,j,l_3,l_4}(0,0,t,-t)\nonumber \\
&&
\times D^{u_1(1-u_2)}_{l_4,l_5}(t)\; 
\Gamma^{(4)}_{r,s,l_5,l_6}(0,0,t,-t)\;D^{u_1u_2}_{l_6,l_1}(t) \Biggr )
\nonumber \\
&&
+12 {\cal P}_{(i,j,r,s)} 
\Biggl (
\Gamma^{(4)}_{w,i,l_1,l_2}(q,0,t,-q-t)\;D^{1-u_1}_{l_2,l_3}(q+t)\;
\Gamma^{(4)}_{z,j,l_3,l_4}(-q,0,q+t,-t)\nonumber \\
&&
\times D^{u_1(1-u_2)}_{l_4,l_5}(t)\;
\Gamma^{(4)}_{r,s,l_5,l_6}(0,0,t,-t)\;D^{u_1u_2}_{l_6,l_1}(t)\;
\nonumber \\
&&
+\Gamma^{(4)}_{w,i,l_1,l_2}(q,0,t,-q-t)\;D^{1-u_1}_{l_2,l_3}(q+t)\;
\Gamma^{(4)}_{r,s,l_3,l_4}(0,0,q+t,-q-t)\nonumber \\
&&
\times D^{u_1(1-u_2)}_{l_4,l_5}(q+t)\;
\Gamma^{(4)}_{z,j,l_5,l_6}(-q,0,q+t,-t)\;
D^{u_1u_2}_{l_6,l_1}(t)\;
\nonumber \\
&&
+
\Gamma^{(4)}_{r,s,l_1,l_2}(0,0,t,-t)\;D^{1-u_1}_{l_2,l_3}(t)\;
\Gamma^{(4)}_{w,i,l_3,l_4}(q,0,t,-q-t)\nonumber \\
&&
\times D^{u_1(1-u_2)}_{l_4,l_5}(q+t)\;
\Gamma^{(4)}_{z,j,l_5,l_6}(-q,0,q+t,-t)\;
D^{u_1u_2}_{l_6,l_1}(t)\;\Biggr )
\nonumber \\
&&
+12 {\cal P}_{(i,j,r,s)} \Biggl ( (w,q) \leftrightarrow (z,-q) \Biggr )
\Biggr \rbrack 
\Biggr \rbrace  +H.O.T.  
\eea
where the sum over repeated internal space indices is understood
and,  for simplicity, the dependence of the 1PI functions on the scale $k$ is 
not indicated explicitly and the following notations have been used
\be
\label{propa}
D^{y}_{i,j}(p)= {\rm exp}\left (-\frac{\Gamma^{(2)}_{i,j}(p,-p) }{k^2} y\right )
\ee
\be
{\rm Tr_t}= \int \frac{ {\rm d}^4 t}{(2\pi)^4}
\ee

In Eq. (\ref{gamma6}) the symbol ${\cal P}_{(i,j,r,s)}$ indicates the sum over the permutations of the indices 
$ i,j,r,s$ but excludes the permutations of the indices belonging to the same function $\Gamma^{(4)}$
(which would correspond to a double counting). Namely the first  ${\cal P}_{(i,j,r,s)}$ in  Eq. (\ref{gamma6}) 
corresponds to six possible permutations, the second one corresponds to twelve  possible permutations,
the third one is just like the second one but with both index $w$ and momentum $q$ simultaneously 
exchanged with $z$ and $-q$. Furthermore in Eq. (\ref{gamma6}) there is also a contribution of  
higher order terms, proportional to  $\Gamma^{(6)}$ and $\Gamma^{(8)}$, indicated as $H.O.T.$ 
which is negligible for the determination of  the two loop  $\beta$-function.
Finally we note that  in Eqs. (\ref{gamma4},\ref{gamma6}) only particular configurations of the momenta of  
$\Gamma^{(4)}$ and $\Gamma^{(6)}$ are considered. In fact these are the configurations which, as we shall see
below, are relevant for the two loop $\beta$-function and anomalous dimension.

The perturbative computation is performed by expressing each  $\Gamma^{(n)}$ as a series in powers of the coupling 
$\lambda$ and reading  from Eq. (\ref{clact}) the lowest non-vanishing terms (which are indicated with a hat)
\be 
\label{lo2}
\widehat \Gamma^{(2)}_{i,j}(q,-q)=q^2 \delta_{ij}
\ee 
\be
\label{lo4}
\widehat \Gamma^{(4)}_{i,j,r,s}(q_1, q_2, q_3, -q_1-q_2-q_3)=\frac{\lambda}{3}\;\Delta(i,j,r,s)
\ee
where $\delta_{ij}$ is the Kronecker delta and 
\be
\label{dema}
\Delta(i,j,r,s)=(\delta_{ij}\delta_{rs}+\delta_{ir}\delta_{js}+\delta_{is}\delta_{jr})
\ee 
When Eqs. (\ref{lo2},\ref{lo4}) are inserted in the right hand side of the flow equations, one can recover the 
first order corrections to the  $n$-point functions, which in turn can be put again in the flow equations 
to get the second order and so on.  By collecting in each  $n$-point function all contributions with  
a fixed power of the coupling constant one obtains the various orders of the perturbative expansion.

We shall first consider the anomalous dimension $\eta$, defined as 
\be
\label{eta}
\eta \; \delta_{ij}= - k\;{\partial   \over \partial k } \left ( \left \lbrack{\partial  
\Gamma^{(2)}_{i,j} (q,-q)\over \partial q^2 }\right
\rbrack_{q=0}\right )
\ee
Clearly the lowest order in Eq. (\ref{lo2}) does not give any contribution to $\eta$ and it is necessary to 
consider the first correction to $\Gamma^{(2)}$ in Eq. (\ref{gamma2}) which is obtained by replacing 
$\Gamma^{(4)}$ in the right hand side of Eq. (\ref{gamma2}) with its lowest order displayed in Eq. (\ref{lo4}). 
However this correction to $\Gamma^{(2)}$ has no dependence on the external momentum of the 
two-point function and then it does not contribute to $\eta$.  The first non-vanishing contribution 
is obtained by considering the first correction to $\Gamma^{(4)}$ in Eq. (\ref{gamma2}). 
This correction is  to be computed from  Eq. (\ref{gamma4}) and for this purpose the required configuration
of the external momenta of $\Gamma^{(4)}$ is just the one considered in Eq. (\ref{gamma4}).
The lowest contribution coming from Eq. (\ref{gamma4}) is of order $O(\lambda^2)$.
(Note that in Eq. (\ref{gamma2}) there is another contribution of order $O(\lambda^2)$ coming from the 
$O(\lambda)$ correction to $\Gamma^{(2)}$ in the exponential but it is independent of the external 
momentum and therefore irrelevant for $\eta$).

Our next step  for the determination of  $\eta$ is the computation of the $O(\lambda^2)$ 
terms in Eq. (\ref{gamma4}) which yield an external momentum dependence when inserted in 
Eqs. (\ref{gamma2}) and  (\ref{eta}). This means that, of the three terms in Eq. (\ref{gamma4})
quadratic in $\Gamma^{(4)}$, the last one must be neglected. Furthermore,
the term proportional to $\Gamma^{(6)}$ must be discarded because it is 
of order $O(\lambda^3)$ as it is evident from  Eq. (\ref{gamma6}). 
Then we replace the lowest order terms  (\ref{lo2}) and (\ref{lo4}) 
in the remaining part of   Eq. (\ref{gamma4})
and then insert this in Eq. (\ref{gamma2}). From this procedure we do not get the full  $\Gamma^{(2)}$ 
to the order $O(\lambda^2)$ but only the relevant part for determining the anomalous dimension,
and we indicate it as $\Gamma^{(2)\eta}$:
\bea
\label{funda}
&& 
k\;{\partial \Gamma^{(2)\eta}_{i,j}(q,-q) \over \partial k }=-\delta_{ij}\frac{(N+2)\lambda^2}{3k^2}\nonumber \\
&&\times {\rm Tr_p} 
\; \Biggl \lbrace  {\rm exp}\left (-\frac{p^2}{k^2} \right ) \;
\int_\Lambda^k \frac{ {\rm d} x}{x^5}
\; {\rm Tr_t} \Biggl \lbrace   \int_0^1 {\rm d}u_1\;{
\rm exp}\left (-\frac{t^2}{x^2} (1-u_1)\right )\nonumber \\
&&
\times\left \lbrack
{\rm exp}\left (-\frac{(t+p+q)^2}{x^2} u_1\right ) +
{\rm exp}\left (-\frac{(t+p-q)^2}{x^2} u_1\right ) \right \rbrack  \Biggr \rbrace 
\Biggr \rbrace 
\eea

In Eq. (\ref{funda}) the factor $(N+2)/3$ comes from the various sums over the internal indices 
and the integral on the variable $x$ corresponds to the integration of Eq. (\ref{gamma4}) to get  
$\Gamma^{(4)}$ at the scale $k$ (note that we are omitting the boundary 
value of  $\Gamma^{(4)}$ at the Ultraviolet (UV) extremum $\Lambda$ because this bare 
value  does not depend on the momentum $q$ and $\Lambda$ will be sent to infinity in this 
calculation because this does not 
introduce any  UV divergences).

At this point the consistent determination of $\eta$ through Eq. (\ref{eta}) 
would require a direct resolution of the integrals in  Eq. (\ref{funda}). 
This can be easily performed but it does not lead to the correct two loop 
coefficient of the squared coupling $(\lambda/(16 \pi^2))^2$. For instance 
in the $N=1$ case instead of the correct  $1/6$ coefficient, the value 
$0.1278$ is obtained.
However it is still possible to recover the two loop anomalous dimension 
from Eq. (\ref{funda}) in the following way.
Let us consider the right hand side of (\ref{funda}). The integrand of the trace
${\rm Tr}_p$  has two factors: the vertex $\Gamma^{(4)}$ obtained from the corresponding
flow equation and the exponential factor $ {\rm exp}\left (-p^2/k^2 \right )$.
Before performing the integrals we consider the limit $k\to 0$ of these two separate terms in 
order to get the full contribution, up to $k=0$, of the four-point function to $\eta$,
and we note that the exponential term in this  limit is singular 
and becomes a $\delta$-function according to Eq. (\ref{delt}) with the position  $k=\epsilon$.
When we replace the exponential with four  $\delta$-functions, one for each component
of the momentum, according to Eq. (\ref{delt}) we get
\bea
\label{fundad}
&& 
k\;{\partial \Gamma^{(2)\eta}_{i,j}(q,-q) \over \partial k }=-\delta_{ij}\frac{(N+2)\pi^2 \lambda^2 }{3}\nonumber \\
&&\times {\rm Tr_p} 
\; \Biggl \lbrace  \delta^4(p) 
k^2 \int_\Lambda^k \frac{ {\rm d} x}{x^5}
\; {\rm Tr_t} \Biggl \lbrace   \int_0^1 {\rm d}u_1
\;{\rm exp}\left (-\frac{t^2}{x^2} (1-u_1)\right )\nonumber \\
&&
\times\left \lbrack 
{\rm exp}\left (-\frac{(t+q)^2}{x^2} u_1\right ) +
{\rm exp}\left (-\frac{(t-q)^2}{x^2} u_1\right ) \right \rbrack  \Biggr \rbrace 
\Biggr \rbrace 
\eea
In Eq. (\ref{fundad}) the limit $k\to 0$ of the exponential has already been performed but 
we still have to consider the same limit  in  the remaining $k$ dependent part of the equation
and it is important to remark that the factor $k^2$ which appears explicitly 
in the trace in Eq. (\ref{fundad}) cancels against a factor  $(1/k^2)$ coming 
from the $x$  integration, so that the final result  is a finite number. 
The integrals over the momenta $p$ and $t$ are gaussian and they 
are easily performed and after that, according to Eq. (\ref{eta}), one can derive with respect to 
$q^2$ and then evaluate at $q=0$ and finally perform the remaining integrals with 
$\Lambda\to+\infty$.  This procedure yields 
\be
\label{etares}
\eta=\left ( \frac{\lambda}{16\pi^2}\right )^2 \frac{(N+2)}{18} 
\ee
Incidentally we note that the determination of the two loop anomalous dimension 
for the $N=1$ theory from the PTRG flow equation,  by making use of the derivative 
expansion according to a procedure previously introduced in \cite{io2l},
has been already discussed in \cite{maza}.

Let us now turn to the $\beta$-function. Again the correct perturbative procedure 
only provides the one loop $\beta$ and fails at two loop. Nevertheless the 
particular limit  considered  for the anomalous dimension works even in this 
case, as discussed below.
The one loop ($O(\lambda^2)$)  $\beta$-function is easily obtained from  
Eq. (\ref{gamma4}) evaluated at zero external momenta $q=p=0$
and replacing the four-point and two-point functions in the right hand side with their lowest order 
expressions in Eqs. (\ref{lo2}) and  (\ref{lo4}). This yields
\be
\label{beol}
k\;{\partial \Gamma^{(4)}_{i,j,r,s}(0,0,0,0) \over \partial k }=
\frac{(N+8)\lambda^2}{9\;k^4} \Delta(i,j,r,s)\; {\rm Tr_t} 
\; \Biggl \lbrace   \;{\rm exp}\left (-\frac{t^2}{k^2} \right )\; \Biggr \rbrace
\ee

As it has been done for the anomalous dimension, we evaluate the trace in Eq. (\ref{beol}) in the limit 
$k\to 0$ and make use of Eq. (\ref{delt}). The one loop $\beta$-function is then obtained from Eq. (\ref{beol})
by discarding the factor $\Delta(i,j,r,s)/3$ which provides the normalization of the lowest order in Eq. (\ref{lo4}):
\be
\label{b1l}
\beta^{1-loop}=\frac{\lambda^2}{16\pi^2} \frac{(N+8)}{3}
\ee
Incidentally we note that, to this level, the trace in Eq. (\ref{beol}) would have given the same output even 
if it were evaluated without taking the limit  $k\to 0$. However when considering higher order calculations
this property is no longer true.

The $O(\lambda^3)$ or two loop $\beta$-function  is the sum of many contributions.
In fact the $\beta$-function describes the evolution of the renormalized 
four point function $\Gamma^{(4) R}$ with vanishing external momenta 
(note that as in Eq. (\ref{b1l}) the normalization factor  
$\Delta(i,j,r,s)/3$ of the coupling in Eq. (\ref{lo4})
is not included in the definition of the $\beta$-function )
\be
\label{definition}
\beta\;\frac{\Delta(i,j,r,s)}{3}=k\;{\partial \Gamma^{(4) R}_{i,j,r,s}(0,0,0,0) \over \partial k }=
k\;{\partial (Z_0^{-2}\Gamma^{(4) }_{i,j,r,s}(0,0,0,0)) \over \partial k }
\ee
where  $Z_0$ is the field independent part of the wave function renormalization $Z_k$ 
and it is obtained  from the two-point function as
\be
\label{zeta}
Z_0=\left \lbrack{{\partial \Gamma^{(2)}_{i,i} (q,-q)}\over {\partial q^2 }} \right \rbrack_{q=0}
\ee
The index $i$ in the right hand side is not summed and can be fixed arbitrarily due to the $O(N)$ symmetry.
Then it is clear from Eq. (\ref{definition}) that there are two separate terms $\beta_g$ and $\beta_\eta$
\bea
\label{duepezzi}
&&
\beta\;\frac{\Delta(i,j,r,s)}{3}=\left (Z_0^{-2}  \beta_g + \beta_\eta\right )\;\frac{\Delta(i,j,r,s)}{3}
\nonumber \\
&&
= Z_0^{-2} \;k\;{\partial \Gamma^{(4) }_{i,j,r,s}(0,0,0,0)\over \partial k }-2\Gamma^{(4) }_{i,j,r,s}(0,0,0,0)\;
k\;{\partial Z_0 \over \partial k }
\eea
and, since $Z_0=1+O(\lambda^2)$, as it follows from  Eqs. (\ref{lo2}), (\ref{eta}),  (\ref{zeta}),  
we see that, to order $O(\lambda^2)$ the $\beta$-function is simply given by $\beta_g$ which is what has been 
computed in Eq. (\ref{beol}). To order  $O(\lambda^3)$ we have from  Eqs. (\ref{zeta}), (\ref{eta}) and (\ref{etares})
\be
\label{betaeta}
\beta_\eta=\frac{\lambda^3}{(16\pi^2)^2} \frac{(N+2)}{9} 
\ee

We must compute now  $\beta_g$ to order $O(\lambda^3)$ and we notice that its coefficient  $Z_0^{-2}$  in
Eq. (\ref{duepezzi}) can be neglected because it gives  contributions to $\beta$ which are at least 
of order $O(\lambda^4)$.  
Then, looking at the right hand side of Eq. (\ref{gamma4})  we can 
isolate three different contributions to  $\beta_g$.
The first one, which will be indicated as $\beta_1$, 
comes from the six-point function which is $O(\lambda^3)$, as Eq. (\ref{gamma6}) shows;
the second one ($\beta_2$) is originated by the  terms quadratic in  $\Gamma^{(4)}$ when one of these
four-point functions is replaced with its first order expression  ($O(\lambda^2)$) and the other is kept 
to the lowest order; finally the last one ($\beta_3$) comes from the same terms with  both  lowest 
order $\Gamma^{(4)}$ and considering the $O(\lambda)$ correction in the 
two-point functions appearing in the exponential factors.

The term $\beta_1$ requires the computation of $\Gamma^{(6)}$ for the particular external momenta displayed
in Eq. (\ref{gamma6}). Then we insert Eqs. (\ref{lo2}) and (\ref{lo4}) into the right hand side of 
Eq. (\ref{gamma6}) and we also sum over the last two indices ( $w$ and $z$) 
because this is required when $\Gamma^{(6)}$  is inserted into Eq. (\ref{gamma4})
\bea
\label{g6out}
&& 
k\;{\partial \Gamma^{(6)}_{i,j,r,s,w,z}(0,0,0,0,q,-q) \over \partial k }\;\delta_{wz}
\nonumber \\
&&
=-\frac{\lambda^3}{27 k^6}\;\Delta(i,j,r,s)\;{\rm Tr_t}
\;\int_0^1 {\rm d}u \; u 
\Biggl \lbrace
2(N^2+10N+16) \; {\rm exp}\left (-\frac{t^2}{k^2}\right )
\nonumber \\
&&
+8 (5N+22)
{\rm exp}\left (-\frac{(t+q)^2}{k^2}(1-u)\right )
\;{\rm exp}\left (-\frac{t^2}{k^2}u\right ) \Biggr \rbrace
\eea
The output of Eq. (\ref{g6out}) must be used in 
Eq. (\ref{gamma4}) where we have the trace of  $\Gamma^{(6)}$ times an exponential term.
Again, before performing the trace  we take  the limit $k\to 0$ of these two terms 
and the exponential  becomes a  $\delta$-function on the four components 
of the momentum. Then it is straightforward to evaluate all the integrals and take the limit
$\Lambda\to +\infty$ which yields 
\bea
\label{be1}
&& 
\beta_1=-\frac{\lambda^3\pi^2}{9}\; {\rm Tr_p} \Biggl \lbrace
\delta^4(p) \;k^2\;\int^{+\infty}_k \frac{ {\rm d} x}{x^7} \; {\rm Tr_t}
\int_0^1 {\rm d}u \;u
\Biggl \lbrack 
2(N^2+10N+16)\; {\rm exp}\left (-\frac{t^2}{x^2}\right )
\nonumber \\&&
+ 8 (5N+22) \;{\rm exp}\left (-\frac{(t+p)^2}{x^2}(1-u)\right )
\;{\rm exp}\left (-\frac{t^2}{x^2}u\right ) \Biggr \rbrack  \Biggr \rbrace=
\nonumber \\&&
=-\frac{\lambda^3}{(16\pi^2)^2}\frac{(N^2+30 N+104)}{18}
\eea

The procedure  to calculate $\beta_2$ is analogous. 
The first step is to evaluate $\Gamma^{(4)}$ to order  $O(\lambda^2)$ and the second step is to  
insert  the result in Eq. (\ref{gamma4}).  
However in the second step we are interested in the limit $k\to 0$ which corresponds,
as it has been done in the previous cases,  to evaluate the $O(\lambda^2)$ four-point function
with all vanishing external momenta, due to the presence of the $\delta$-function.
Therefore we can directly evaluate the $\Gamma^{(4)}$ to order $O(\lambda^2)$
with all  external momenta put to zero by integrating Eq. (\ref{gamma4}) from $\Lambda$ to $k$.

However we recall that the perturbative four-point function 
requires the introduction of a counterterm to cancel the divergences occurring in the computation.
In fact with the  renormalization condition that $\Gamma^{(4)}$ at zero external momenta is  
equal (apart from the normalization constant) to the 
coupling $\lambda$:  $\Gamma^{(4)}_{i,j,r,s}(0,0,0,0)=\lambda\;\Delta(i,j,r,s)/3$, 
and by comparison with the lowest order prescription in Eq. (\ref{lo4}),
we see that this  counterterm must cancel the  $O(\lambda^2)$ 
contribution to the four-point function with vanishing  external momenta,
when the limits $\Lambda\to +\infty$ and $k\to 0$ are taken.
Therefore in this case there is no contribution to the $\beta$-function, i.e. 
\be
\label{be2}
\beta_2=0
\ee

Finally we consider $\beta_3$ which comes from the  two-point function. We compute  the $O(\lambda)$ correction 
to $\Gamma^{(2)}$ from  Eq. (\ref{gamma2}). We get
\be
\label{g2res}
k\;{\partial \Gamma^{(2)}_{i,j}(q,-q) \over \partial k }
=-\frac{\lambda k^2}{16\pi^2}\frac{(N+2)}{3}\;\delta_{ij}
\ee

As for $\Gamma^{(4)}$ we need a counterterm which is fixed by the renormalization condition 
that the theory is massless. Therefore this counterterm must cancel exactly the term quadratic
in $\Lambda$ coming from the integration of Eq. (\ref{g2res}) so that 
we are left with a mass term proportional to $k^2$  which as required, 
is vanishing in the limit  $k\to 0$. Therefore, once the quadratically divergent
term has been cancelled, $\Gamma^{(2)}$ has the form 
$\Gamma^{(2)}_{i,j}(q,-q)=(q^2-\lambda(N+2) k^2/(6 \cdot 16\pi^2 ))\delta_{ij}$
and the exponential factor in Eq. (\ref{gamma4}) containing  $\Gamma^{(2)}$ 
can be expanded in powers of the coupling 
and the $O(\lambda)$ term  can be used in Eq. (\ref{gamma4}) to get $\beta_3$. 
This computation is straightforward 
\bea
\label{be3}
&& 
\beta_3=\frac{(N+8)\lambda^2}{3 k^4}\; {\rm Tr_p} \Biggl \lbrace \frac{\lambda(N+2) }{6 (16\pi^2)}
{\rm exp}\left ( -\frac{p^2}{k^2} \right ) \Biggr \rbrace
\nonumber \\ &&
=\frac{\lambda^3}{(16\pi^2)^2 }\frac{(N+8)(N+2)}{18}
\eea
where, again,  we have taken the limit $k \to 0$ and 
reduced the exponential to a $\delta$-function and
eventually performed the trace.

Finally we collect  all contributions to $\beta_g=\beta_1+\beta_2+\beta_3$ in Eqs. (\ref{be1},\ref{be2},\ref{be3})
and according to Eqs. (\ref{duepezzi}) and (\ref{betaeta}) we get 

\be
\label{bfin}
\beta^{2-loop}=\beta_\eta+\beta_1+\beta_2+\beta_3=-\frac{\lambda^3}{(16\pi^2)^2}\frac{(3N+14)}{3} 
\ee
Eqs. (\ref{etares}), (\ref{b1l}) and (\ref{bfin}) reproduce the correct perturbative anomalous dimension 
and $\beta$-function which had been already obtained by means of the ERG (e.g. see \cite{pw}). 

Let us briefly comment on these results. The fact that the two loop 
$\beta$-function and anomalous dimension are not obtained from 
the standard perturbative analysis of Eq. (\ref{eq:gamma}) is in agreement
with \cite{lpmain} where it is shown that the PTRG does not reproduce the 
two loop effective action. On the other hand we have seen that there exists
a particular limit in which the flow equation gives the two loop 
$\beta$ and $\eta$. Let us reconsider the crucial step of the 
calculation of $\eta$ (the case of the $\beta$-function is analogous).
We have checked that at any finite value of $k$  Eq. (\ref{funda})
yields a wrong coefficient of $(\lambda/(16 \pi^2))^2$ for $\eta$.
This wrong coefficient does not depend on the scale $k$ which means that 
it stays unchanged in the limit $k\to 0$. The situation is different if one tries 
to compute the right hand side  of Eq. (\ref{funda}) directly at $k=0$
because of the appearance of some singularities.
In fact, in the case considered above,
 the exponential term  ${\rm exp}\left (-\frac{p^2}{k^2} \right )$
in  Eq. (\ref{funda}) shows a singularity that  corresponds to a 
$\delta$-function which, in turn, leads to the correct value of $\eta$.
Therefore by exchanging the operations of taking the limit $k \to 0$
and performing the integrals, one gets different results.
So either one concludes that the results in Eqs. (\ref{etares}) and (\ref{bfin})
are purely accidental  or one has to take them as a hint that the flow 
equation here considered has a peculiar limit $ k \to 0$ which contains 
some correct features of the exact theory.

In principle one could perform the same analysis by starting with the proper time flow equation 
regulated by the $m$-dependent smooth cut-off. It is however very problematical to solve 
the various integrals  in this case both in the limit $k\to 0$ and at finite $k$. It is instead 
quite simple to check that the 
$m$-dependent  flow  equations for the various $n$-point functions become again the 
exponential flow equations considered above when the limit $m\to \infty$ is considered 
and therefore, even in this case, the same results for  $\eta$ and $\beta$ are obtained.

Another remark concerns the addition of a mass term in Eqs. (\ref{clact}) and (\ref{lo2}). 
The effect on the exponential flow equations  is the appearance of a squared mass 
term $M^2$  in the two-point function in Eq. (\ref{propa}), which
can always be disentangled from the momentum dependent part of the two-point function
due to the exponential form of $D^y_{i,j}(p)$. 
Therefore the flow equations will contain multiplicative
factors like ${\rm exp}(-M^2/k^2)$ and it is not difficult to realize that in this case 
the limit $k\to 0$ corresponds to  a vanishing $\beta$-function.
Clearly this is an indication that in the massive case, when $k$ is approaching zero and is much 
smaller than the mass, the flow has stopped and the various running quantities have 
become constant. Then  the approach used here yields the perturbative results 
only if the flow is confined to the critical surface which means that 
the mass counterterm is fixed in such a way that the theory is massless.

\section{Proper time  and exact flow}

In order to have a clearer picture of the differences between the ERG and the PTRG,
in this Section we follow \cite{wetplb}, where a full derivation of the ERG from the 
fundamental path integral is described, and see to which extent it is possible 
to reformulate this argument for the PTRG.
Let us briefly recall the main points presented in \cite{wetplb}. In the following the IR regulator
is indicated as $H(k;x,y)$ instead of $R_k$ because, as we shall see, they have quite different features.
The connected Green function generator $W_k$, which depends
on the two sources $J(x)$ and $H(k;x,y)$, is  (in the following the 
integration over the spatial coordinates will
be indicated with  a center dot)
\be
\label{eq:start}
\exp \left ( {W_k[J,H]}\right ) =\int\!\!{\cal D}\phi\,\exp\left (-S+J\cdot\phi-\frac{1}{2}\phi\cdot H\cdot\phi
\right )
\ee
and, as usual:
\be\label{eq:defder}
\frac{\delta W_k}{\delta J}=\langle\phi\rangle = \varphi\quad
\ee
\be
\frac{\delta^2 W_k}{\delta J\delta J}=\langle\phi\phi\rangle-\varphi\varphi,\quad
\frac{\delta W_k}{\delta H}=-\frac{1}{2}\langle\phi\phi\rangle
\ee
where we have defined 
\be
\langle A\rangle= \exp{\left (-W_k[J,H]\right )} \int
\!\!{\cal D}\phi\,A
\exp(-S+J\cdot\phi-\frac{1}{2}\phi\cdot H\cdot\phi)
\ee
The two sources $J(x)$ and $H(k;x,y)$ (from now on the $x$ and $y$ dependence of $H$ will not be 
explicitly displayed in  order to have a simpler notation)  are independent and we can consider variations of 
$W_k$ with respect to one of them while keeping fixed the other.
Now we  consider the partial Legendre transform of $W_k$ with respect to $J$ 
\be
\label{eq:puccio}
\widetilde\Gamma_k[\varphi,H]= -W_k[J,H]+J\cdot\varphi
\ee
where $\varphi$ has been defined in  Eq. (\ref{eq:defder})  and $J$ in the 
r.h.s. of Eq. (\ref{eq:puccio}) is expressed as 
\be
\label{eq:pippo}
J=J[\varphi,H]
\ee
which is obtained by inverting Eq. (\ref{eq:defder}).

Then, as usual we can compute some functional derivatives
\be
\frac{\delta\widetilde\Gamma_k}{\delta \varphi}=-\frac{\delta W_k}
{\delta J}\cdot \frac{\delta J}{\delta\varphi}+
\varphi\cdot \frac{\delta J}{\delta\varphi}+J=J
\ee
\be
\label{eq:dersec}
\frac{\delta^2\widetilde\Gamma_k}{\delta \varphi\delta \varphi}=\frac{\delta J}{\delta\varphi}=
\left(\frac{\delta^2 W_k}{\delta J\delta J}\right)^{-1}
\ee

The average effective action $\Gamma_k$ is defined as
\be
\label{eq:puccio1}
\Gamma_k[\varphi,H]=\widetilde \Gamma_k[\varphi,H]-\frac{1}{2}\varphi\cdot H\cdot\varphi
\ee
and it is easy to check that the source $J$ which is expressed through Eq. (\ref{eq:pippo}) is to
be considered as a $k$-dependent term, whereas $\varphi$ is an independent variable.
Then, the second functional derivative is
\be
\frac{\delta^2 \Gamma_k}{\delta \varphi\delta \varphi}=
\frac{\delta^2\widetilde \Gamma_k}{\delta \varphi\delta \varphi}-H
\ee
where as stated before the derivation with respect to 
$\varphi$ is done with  $H$  fixed. Recalling that 
$H$ depends on $k$, it follows:
\be
\partial_k\widetilde \Gamma_k=-\partial_k W_k-\frac{\delta W_k}
{\delta J}\cdot \partial_k J+\varphi\cdot \partial_k J
\ee
\be
=-\partial_k W_k=\frac{1}{2}\mathrm{Tr}\Big(\partial_k H\cdot \langle\phi\phi\rangle\Big)
\ee
For the latter equation it must be noticed that the 
source $J$ which is expressed through Eq. (\ref{eq:pippo}) is to
be considered as a $k$-dependent term, whereas $\varphi$ is an independent variable.
Then for $\Gamma_k$ we get:
\begin{displaymath}
\partial_k \Gamma_k=\partial_k\widetilde \Gamma_k-
\frac{1}{2}\varphi\cdot\partial_k H\cdot\varphi=
\frac{1}{2}\mathrm{Tr}\left\{\partial_k H\cdot[\langle\phi\phi\rangle-\varphi\varphi]\right\}
\end{displaymath}
\be
=\frac{1}{2}\mathrm{Tr}\left\{\partial_k H\cdot\left(\frac{\delta^2 \Gamma_k}
{\delta \varphi\delta \varphi}+H\right)^{-1}\right\}.
\ee
This is the flow equation for $\Gamma_k$. Moreover the specific requirement 
that $H(k=0)=0$ implies that the  functionals $W_k$ and $\Gamma_k$
become equal to the connected Green function generator and to the 
effective action respectively, as can be directly checked from their definitions 
in Eqs.  (\ref{eq:start}),  and (\ref{eq:puccio1}).

Furthermore, if the regulator, evaluated at the UV cut-off 
$H(k=\Lambda)$ for very large  $\Lambda$, is actually divergent, then at 
that scale the average effective action reduces to the original action
\be\label{eq:laminf}
\Gamma_\Lambda=S
\ee
To prove Eq. (\ref{eq:laminf}) let us consider the following shift  
in the functional integral in Eq. (\ref{eq:start})
\begin{displaymath}
\phi=\varphi+\phi'
\end{displaymath}
where $\varphi$ is a background field which we choose equal to the expectation value of the field defined in 
Eq. (\ref{eq:defder})   
\begin{displaymath}
\exp(W_k)=\int\!\!{\cal D}\phi' \exp\left(-S[\phi'+\varphi]+
J\cdot(\phi'+\varphi)-\frac{1}{2}(\phi'+\varphi)\cdot
H\cdot(\phi'+\varphi)\right)
\end{displaymath}
\begin{displaymath}
=\int\!\! {\cal D}\phi' \exp\left(-S[\phi'+\varphi]+J\cdot\phi'-\frac{1}{2}\phi'\cdot H\cdot
\phi'\right)
\end{displaymath}
\be
\times \exp\left(J\cdot\varphi-\frac{1}{2}\varphi\cdot H\cdot\varphi\right)\exp\left ( -\frac{1}{2}\left(
\phi'\cdot H\cdot\varphi+\varphi\cdot H\cdot\phi'\right)\right )
\ee

By deriving  (\ref{eq:puccio1}) one gets
\be \label{eq:degamk}
\frac{\delta \Gamma_k}{\delta\varphi}=J-\frac{1}{2}H\cdot\varphi-\frac{1}{2}
\varphi\cdot H
\ee
and therefore, by inserting   (\ref{eq:puccio1}) and  (\ref{eq:degamk}) one gets
\be
\label {eq:semicl}
\exp(-\Gamma_k)=\int\!\!{\cal D}\phi' \exp\left(-S[\phi'+\varphi]+
\frac{\delta \Gamma_k}{\delta\varphi}\cdot\phi'-\frac{1}{2}\phi'\cdot H\cdot\phi'\right)
\ee
If $H$ diverges in the limit  $k=\Lambda\to \infty$,
the term $\exp(\phi'\cdot H\cdot\phi'/2)$ behaves like a functional delta
$\delta[\phi']$ in that limit, and therefore Eq. (\ref{eq:laminf}) follows from  Eq. (\ref{eq:semicl}).

Let us now turn to the proper time flow equation and reconsider again the problem 
with a different input.
As a first remark we note that in Eq.  (\ref{eq:defder}) $\varphi$ is defined by a functional derivative of 
$W_k$  with respect to $J$ and keeping $H$ fixed.  We keep the same definition of  $\varphi$,
but then we evaluate $W_k$ and $\delta W_k / \delta J$ for a particular 
source $H$, namely  
\be
\label{eq:hbar}
H=\overline{H}(\varphi,k;x,y)
\ee
whose explicit form will be given later ( as before the $x$ and $y$ dependence in $\overline H$ will not be displayed). 
To keep track of this important difference with the previous case we shall use the new notation 
$\overline W$, $\overline \Gamma$ $\widetilde {\overline \Gamma}$  (for simplicity the dependence 
on the scale $k$ is omitted), instead of $W_k$, $\Gamma_k$ 
and $\widetilde {\Gamma}_k$.
Obviously with this choice Eq.  (\ref{eq:defder})  is no longer  a simple definition of
$\varphi$ and it has become an equation which defines $\varphi$ only implicitly. An explicit 
expression of $\varphi$ would be obtained only solving the equation.  Nevertheless we define  
 $\varphi$ as the solution of this equation, that is  $\varphi(x)=\varphi[J,\overline{H}(\varphi,k)]$
and $J$ can be obtained by inverting the latter  
\be
\label{eq:gei}
J=J[\varphi,\overline H]
\ee
Moreover, as in the derivation considered before, $J$ depends on $k$
through its dependence on $\overline H$ in Eq. (\ref{eq:gei}).

We can go on and define as before
\be
\widetilde {\overline{\Gamma}}[\varphi,\overline{H}]= -\overline{W}[J,\overline{H}]+J\cdot\varphi
\ee
\be
\label {eq:defi}
{\overline{\Gamma}}=\widetilde{\overline{\Gamma}}-\frac{1}{2}\varphi\cdot\overline{H}\cdot\varphi.
\ee
Again in these two equations $J$ is to be expressed through  Eq. (\ref{eq:gei}).

It is clear that both $\widetilde{\overline{\Gamma}}$ and ${\overline{\Gamma}}$  due to Eq. (\ref{eq:hbar})
depend only on $\varphi$ and $k$. Then it must be remarked that many equations  
which hold for a simple $\varphi$-independent source $H$, here are no longer valid.
For instance if we evaluate $\delta\widetilde{\overline{\Gamma}}/\delta\varphi$ (we intend here the full
variation of $\widetilde{\overline{\Gamma}}$ with respect to $\varphi$, including the implicit dependence
through $\overline H$) we get
\be
\frac{\delta\widetilde{\overline{\Gamma}}}{\delta\varphi}=-\frac{\delta\overline{W}}
{\delta J}\cdot\frac{\delta J}{\delta\varphi}-
\frac{\delta\overline{W}}{\delta \overline H}\cdot\frac{\delta\overline{H}}{\delta\varphi}+
\frac{\delta J}{\delta\varphi}\cdot \varphi+J=
J-\frac{\delta\overline{W}}{\delta \overline H}\cdot\frac{\delta\overline{H}}{\delta\varphi}
\ee
Therefore it is no longer possible to recover  the simple relation in Eq. (\ref{eq:dersec})
between 
$\delta^2\widetilde{\overline{\Gamma}}/(\delta\varphi\delta\varphi)$ and  
$\delta^2\overline{W}/(\delta J\delta J)$ (which, as already discussed 
when we introduced $\overline H$, is the second functional derivative of $\overline{W}$
with respect to $J$ but keeping $\overline H$ fixed).
 
With this caveat we derive the flow equation 
\be
\partial_k\widetilde{\overline{\Gamma}}
=\frac{1}{2}\mathrm{Tr}\left(\partial_k \overline{H}\cdot\langle\phi\phi\rangle\right)
\ee
\be\label{eq:hfix}
\partial_k\overline{\Gamma}=\frac{1}{2}\mathrm{Tr}\left[
\partial_k \overline{H} \cdot \left(\frac{\delta^2\overline W}
{\delta J\delta J}\right)\right].
\ee
At this point the proper time flow equation is easily obtained  from Eq. (\ref{eq:hfix}). In fact
now we choose the explicit form of the source $\overline H$ to be  
\be
\label{eq:defH}
\overline{H}(x,y)=2{\displaystyle\int_0^k\!\!\mathrm{d}q\,\left\{\left[
\overline{W}_q^{''}(x,x')\right]^{-1}\cdot
\frac{1}{q}\exp\left (-\frac{1}{Z_q q^2}\overline{\Gamma}_q^{''}(x',y)\right ) \right \} }\ee
where we have used the following notations 
\be
\label{wino}
\left[\overline{W}_q^{''}(x,x')\right]^{-1}=\left[\left(\frac{\delta^2\overline{W}}
{\delta J(x)\delta J(x')}\right)_{J=J[\varphi,\overline H(\varphi,q)]}
\right]^{-1}
\ee
and
\be
\label{wano}
\overline{\Gamma}_q^{''}(x',y)= \left(\frac
{\delta^2\overline{\Gamma}}{\delta\varphi(x')
\delta\varphi(y)}\right)_{\overline H=\overline H(\varphi,q)}
\ee
and the dot in Eq. (\ref{eq:defH}) indicates integration with respect to $x'$. For the sake of clarity
the dependence on the scale $q$ is indicated explicitly in Eqs. (\ref{wino}) and (\ref{wano}).
Note again that in Eq. (\ref{eq:defH}) $\left[\overline{W}_q^{''}(x,x')\right]^{-1}$
is kept explicitly because it cannot be simply replaced with the second derivative of 
$\overline{\Gamma}$. By replacing Eq. (\ref{eq:defH}) into Eq. (\ref{eq:hfix}) one gets 
\be\label{eq:hkr}
k\partial_k\overline{\Gamma}=
\mathrm{Tr}\left[ \exp{\left (-\frac{1}{Z_k k^2}\overline{\Gamma}_k ^{''}\right)}\right]
\ee
which is precisely the flow equation in Eq. (\ref{eq:gamma}).

Unfortunately in the procedure outlined above the explicit  form of 
$\overline H$ is always hidden and, in general, very difficult to determine 
because it requires one to solve Eq. (\ref{eq:defH}). However, 
at least for a simple example,
it is possible  to show how the procedure works.
We consider here, as an application,
the free theory  which can be fully solved. We first remind one  that for 
the free theory the exact computation of the path integral leads 
to the following relation between the classical and the effective action
\be
\label{eq:relbas}
\Gamma=S +\frac{1}{2} {\rm Tr} \left ( \log (S'' ) \right )
\ee
where the second derivative of the action is field independent and in the momentum space  
reads $S''=p^2+M^2$.
The divergent term $(1/2){\rm Tr} (\log S'')$ is usually 
cancelled by a normalization factor in  the path integral  and the conclusion
is that $\Gamma=S $. Let us go back to the proper time flow equation which is solved 
straightforwardly 
\be
\label{eq:solflo}
\Gamma_k - S={\rm Tr} \left (\int^k_\Lambda \frac{ {\rm d}x}{x}\; e^{-S''/x^2} \right )
\ee
We have used the boundary condition that at $k=\Lambda$ the running 
effective action  $\Gamma_\Lambda$ is equal to $S$ and it is easy to realize that in 
the right hand side of Eq. (\ref{eq:solflo}) the second derivative of the running 
effective action is $S''$ because the flow, due to the boundary condition, 
modifies $\Gamma_k$ only by a field independent 
term which cannot affect $\Gamma_k''$ and therefore $\Gamma_k''=S''$. 
Let us  consider the IR limit in  Eq. (\ref{eq:solflo})  putting $k=0$
and, by identifying the running action at $k=0$ with the effective action $\Gamma$
we get 
\be\label{eq:esgam}
\Gamma=S -\frac{1}{2} {\rm Tr} \left (\int^\infty_{1/\Lambda^2} \frac{ {\rm d}\tau}{\tau}\; e^{-\tau S''} \right )
\ee
which is essentially equivalent to Eq.  (\ref{eq:relbas}) and this time the 
cancellation of the infinite constant (when the UV cut-off $\Lambda$ is sent to infinity
the integral is divergent) in Eq.  (\ref{eq:esgam}) is obtained 
by simply modifying the boundary condition of the flow equation with the 
addition of the proper counterterm  to the classical action at the 
scale $\Lambda$. Once $\Gamma_k$ is determined, 
we follow the same procedure used to determine Eq. (\ref{eq:propinv}) and recalling that $\Gamma_k''=S''$
for this particular problem, we get 
\be
\label{eq:vdmu}
\left[\overline{W}_{k}^{''}\right]^{-1} =S'' + \overline H_k
\ee
 where we have explicitly indicated the scale dependence of  $\overline H$.
Now we have all the ingredients to solve Eq. (\ref{eq:defH}) which is  equivalent to 
the differential equation
\be
\label{eq:deph}
k\;{\partial \overline H_k \over \partial k }=
 2 \left \lbrack \left (S'' + \overline H_k \right )
{\rm exp}\left  (-\frac{S''}{k^2}
\right )\right \rbrack
\ee
together with the condition $\overline H_{k=0}=0$. 
Thus we finally get the solution
\be
\overline H_k=S'' \left \lbrack \exp \left ( \int^\infty_{1/k^2} \frac{ {\rm d}\tau}{\tau}\; e^{-\tau S''}  \right )-1
\right \rbrack =
S'' \left \lbrack \exp \left ( \Gamma\left (0,\frac{S''}{k^2}\right ) \right ) -1\right \rbrack
\ee
where we have expressed the integral in terms of the incomplete Gamma function $\Gamma(a,b)$. 
Then, for the non-interacting theory the solution $\overline H$ has a simple field independent form 
which was indicated simply as $H$ at the beginning of this Section. In this case, as discussed before,  
$\overline H$ has exactly the same role of the  regulators $R_k$ employed for the 
Exact Renormalization Group flow and therefore the same interpretation of smooth
cut-off of the IR modes (clearly even the flow equation is of the same form, as can be easily 
checked). 

Let us come back to the comparison with the ERG.
Clearly Eq. (\ref{eq:defH})  is not a simple definition of the regulator, but it is rather an
implicit  equation for $\overline H$ whose resolution is in practice equivalent to solve 
Eq. (\ref{eq:hkr}). This poses a serious limitation on our procedure and in particular 
on the possibility of showing that $\overline{\Gamma}$ interpolates between 
the bare action in the UV and the full effective action in the IR.
In fact according to the derivation outlined above,  $\overline{\Gamma}$
is practically defined through the flow in  Eq. (\ref{eq:hkr}) and $\overline H$
through Eq. (\ref{eq:defH}). The latter can be put in the form of a differential 
equation for $\overline H$ by deriving both sides with respect to the scale $k$.
Then we are able to choose the boundary condition (BC) for $\overline{\Gamma}$
in the UV and determine, by means of the flow equation, the form of 
 $\overline{\Gamma}$ in the IR region. In principle it should not be possible to 
take the BC in the IR and determine the action in the UV but in practice it is 
still possible, provided one puts sufficient constraints on the most general form of
 $\overline{\Gamma}$. Since $\overline H$ is related to  $\overline{\Gamma}$
through Eq.  (\ref{eq:defi}), the BC chosen for $\overline{\Gamma}$ is related to 
the BC of the differential equation for $\overline H$ . 

However, here comes 
the first important difference with respect to the ERG where the explicit form 
of $H$ is given as an input.  Here the explicit form of $\overline H$ is given  by the 
solution of   Eq. (\ref{eq:defH}), which means that we do not have the freedom to 
take for instance in the UV region an $\overline H$ which diverges to get the 
condition $\overline{\Gamma}=S$ at the UV scale $\Lambda$, as in the case of 
the ERG (see Eq. (\ref{eq:laminf})). The only possibility left  is to check that 
if we assume that at the scale $\Lambda$ the relation $\overline{\Gamma}=S$
holds, then Eq. (\ref{eq:defH}) admits a divergent $\overline H$ as a solution.
In this case we can conclude that the two BC $\overline{\Gamma}=S$ and 
infinite $\overline H$  in the UV are consistent.

If one is interested in considering the BC in the IR instead of the UV, the same 
problem appears. For the ERG the cutoff $H$ vanishes at $k=0$ and this 
condition insures that the running action tends to the effective action
when $k\to 0$.  In our case, since the form of $\overline H$ is not explicitly 
available, we can choose the full effective action as BC for  
$\overline{\Gamma}$ at $k=0$, insert it into Eq. (\ref{eq:defH}) and check 
whether this equation consistently admits the condition  $\overline H=0$
at $k=0$ or not.

Actually we are able to check the consistency of these BC for  
$\overline{\Gamma}$ and $\overline H$ both in the UV and in the IR.
Let us first consider the UV case.
We note that if $\overline H$ diverges for $k=\Lambda$ and
$\Lambda\to \infty$ then, by  following the same steps discussed 
above for $H$, we find that  Eq. (\ref{eq:laminf}) holds even in this case.  
From Eq. (\ref{eq:laminf}) we can also compute the second derivative of 
$\overline{W}_q$ which, by definition, is given by $ \overline{W}_q^{''}=\delta \varphi / \delta J$. 
In fact if we use Eq. (\ref{eq:laminf}) to replace $\overline{\Gamma}$ in Eq. (\ref{eq:defi})
we get an expression for $\overline{W}_q$ which can be derived with respect to the source $J$,
keeping $\overline H$ fixed. By  deriving once and recalling that this derivative must be equal 
to the field $\varphi$ by definition, one gets
\be\label{eq:fazero}
\frac{\delta S}{\delta \varphi}=J- \varphi \cdot \overline H
\ee
By deriving once again Eq. (\ref{eq:fazero}) with respect to  $J$ with $\overline H$ fixed we 
recover the desired expression for  $\overline{W}_q^{''}$ in the UV limit:
\be\label{eq:propinv}
\left \lbrack \overline{W}_q^{''}\right \rbrack^{-1}
=S''+\overline H
\ee
So far we have seen that if  $\overline H\to \infty$  when $k=\Lambda$ and $\Lambda\to \infty$ 
then Eqs. (\ref{eq:laminf}) and (\ref{eq:propinv}) hold. Now we can check that these 
relations are consistent with Eq. (\ref{eq:defH}).  In the UV limit the upper extremum of the 
integral in  the right hand side of Eq. (\ref{eq:defH}) tends to infinity and we analyze 
the region of  very large values of $q$
where the exponential is substantially close to one and the relevant term in  
$\left \lbrack \overline{W}_q^{''}\right \rbrack^{-1}$
is $\overline H$.
Then it is easy to realize that Eq. (\ref{eq:defH}) in the 
limit $k=\Lambda\to \infty$  is consistent with 
the divergent solution  $\overline H\propto \Lambda^2$.

Let us consider the IR case. At $k=0$ 
the range of integration
of the variable $q$ in the right hand side of Eq. (\ref{eq:defH})
vanishes but, before concluding that $\overline H(\varphi,k=0)=0$
one has to verify that the integrand is not singular at $k=0$.
We note that if $\overline H(\varphi,k=0)=0$ then,
by using the same steps as in the case 
of the field independent regulator $H$ 
one gets that  $\overline{\Gamma}$ at $k=0$ becomes 
the effective action and $\overline W$ the connected 
Green function generator.
Moreover, at least for a scalar field, the effective action is 
convex and therefore $\overline{\Gamma}^{''}$ is non-negative.
As a consequence, putting $\overline H(\varphi,k=0)=0$ in the 
right hand side of  Eq. (\ref{eq:defH}) means that the 
exponential has a safe limit when $k\to 0$ and  
$\left[\overline{W}_{q=0}^{''}(x,x')\right]^{-1}$ which becomes 
the inverse propagator is also non-singular and 
therefore no singularities appear in the integrand at $k=0$.
In conclusion  $\overline H(\varphi,k=0)=0$ is consistent with 
Eq. (\ref{eq:defH}).
Incidentally in the previous Section the limit $k\to 0$ of the flow equations
 has been already encountered, at least within the  perturbative expansion.
There, since the two-point function was evaluated to the lowest order, we had 
terms like $\exp(-p^2/k^2)$ which, in the limit $k\to 0$ were treated 
as $\delta$-functions for  the momentum $p$. Here in Eq.  (\ref{eq:defH}), the 
same problem can appear for those particular cases in 
which  $\overline{\Gamma}_{q}^{''}$ is vanishing. Then again, 
in the limit $k\to 0$  (and therefore $q\to 0$) the exponential should be treated 
as a $\delta$-function. However no singularity affects Eq.  (\ref{eq:defH}) because 
of the other factor in the trace, $\left[\overline{W}_{q}^{''}(x,x')\right]^{-1}$, which, 
in the limit $k\to 0$  ( $q\to 0$)  becomes exactly  $\overline{\Gamma}_{q=0}^{''}$.
Therefore the argument of the trace has the form $x\delta(x)$ which is not singular. 

So far we have derived the flow in  Eq. (\ref{eq:hkr}) for $\overline{\Gamma}$
by means of a particular cut-off $\overline H$ defined by Eq. (\ref{eq:defH})
and we have checked that we can consistently take as BC of the flow 
either an infinite $\overline H$ with $\overline{\Gamma}=S$ in the UV or 
$\overline H=0$ with $\overline{\Gamma}$ equal to the effective action
in the IR. Of course we do not have the freedom to consider both BC
in the UV and the IR simultaneously. This clearly indicates 
the second important difference with the ERG case where there is a 
suitable cut-off which simultaneously fulfills both conditions,
and it  represents the main limit of this analysis which,
due to the missing link between UV and IR,  cannot be used to  
conclude that the flow in Eq. (\ref{eq:hkr}) is exact.

\section{Conclusions}

We conclude with some comments on the results obtained above.
We have reformulated and modified the derivation of the ERG flow for the case of the PTRG flow.
Since the latter does not belong to the same class of the former, the price we had to pay is 
that, instead of introducing a previously defined regulator as it happens for  $R_k$ in the case 
of the ERG, the regulator $\overline H$ comes out as the solution of a very involved equation.
We have seen that in the simplest case, which corresponds to the free theory,  $\overline H$
can be derived explicitly, it is field independent and it is practically equivalent to a regulator 
$R_k$, but in the interacting case $\overline H$ is field dependent and it is extremely difficult
to determine its form. 
Due to this difficulty, the characteristic properties of $R_k$ in the UV and in the IR,
which insure the exactness of the ERG, cannot be extended to $\overline H$ and the PTRG.

On the other hand we have seen that for the latter flow it is possible to consistently 
choose for the running action  $\overline{\Gamma}$ the BC 
$\overline{\Gamma}=S$ in the UV region or, 
alternatively,  take $\overline{\Gamma}$ equal 
to the effective action in the IR region at $k=0$. 
Once one of these BC is chosen one should be able 
in principle to determine the running action 
at each value of the scale $k$ by solving the flow equation.
This point could be employed, at least in some specific  example
where some informations on the cut-off $\overline H$ can be derived,
 for instance to study the deviation of the PTRG from the classical action 
when one chooses the full effective action as an IR BC.

A different kind of indication has been obtained from the perturbative analysis in Section 2.
As already noticed, in \cite{lpmain} it is discussed that the 
diagrammatic structure beyond one loop is lost in the PTRG flow. This means that in a loop expansion,
starting at two loop one does not recover the correct weights of the various diagrams. The origin of 
this problem can be understood if one looks at Eq. (\ref{eq:gamma}) and expands the exponential 
truncating to the linear term in $\Gamma''/k^2$. Then  one can easily recognize 
that this truncation is equal to the analogous truncation of the expansion of the ERG equation 
where the cut-off $R_k=k^2$ is used. So only in the region where  $\Gamma''<<k^2$  the two flow 
equations have  similar features, but since the trace in the right hand side of the equation involves 
momenta at any scale, there are always regions where the mentioned expansion does 
not hold and consequently the  two flows have different structures. 
In  Section 2 we have checked that the standard perturbative analysis of the flow equation 
leads to uncorrect results at two loop for the $\beta$-function and the anomalous dimension.
It is still possible to extract the two loop results if one considers the limit $k \to 0$ 
before performing all integrals on the momenta, which signals the peculiarity of the 
point $k=0$ for the flow in  Eq. (\ref{eq:gamma}).

The author acknowledges A. Bonanno and M. Reuter for many helpful discussions and suggestions.


\begin{thebibliography}{12}
\bibitem{kad} 
K.G. Wilson, Phys. Rev. {\bf B4}, 3174 and 3184, (1971);
K.G. Wilson and M.E. Fisher, Phys. Rev. Lett. {\bf 28}, 240, (1972);
K.G. Wilson and J. Kogut, Phys. Rep. {\bf 12}, 75, (1974).
\bibitem{polch} J. Polchinski, Nucl. Phys. {\bf B 231}, 269 (1984). 
\bibitem{wet1}
C. Wetterich, Nucl. Phys. {\bf B352}, 529, (1991); 
Z. Phys. {\bf C57}, 451, (1993); {\bf C60} 461, (1993); 
\bibitem{wetplb}C. Wetterich, Phys Lett. {\bf B301}, 90, (1993).
\bibitem{bdm} M. Bonini, M. D' Attanasio and G. Marchesini,  Nucl. Phys.  {\bf B409}, 441, (1993).
\bibitem{ellw} U. Ellwanger, Z. Phys. {\bf C62}, 503, (1994).
\bibitem{mor1} T. Morris, Int. J. Mod. Phys. {\bf A 9}, 2411, (1994);
T. Morris, Phys. Lett. {\bf B329}, 241, (1994).
\bibitem{berg} J. Berges, N. Tetradis and  C. Wetterich,
{\it Nonperturbative renormalization group flow in quantum field 
theory and statistical physics}, Preprint: MIT-CTP-2980, HD-THEP-00-26, 
May 2000  and hep-ph/0005122, submitted to Phys.Rep.
\bibitem{bagn} C. Bagnuls and C. Bervillier, Phys. Rept. {\bf 348}, 91, (2001). 
\bibitem{ole} M. Oleszczuk, Z. Phys. {\bf C64}, 533, (1994).
\bibitem{flo} R. Floreanini and R. Percacci, Phys. Lett. {\bf B356}, 205, (1995). 
\bibitem{sen1} S.-B. Liao,  Phys. Rev. {\bf D53}, 2020, (1996); Phys. Rev. {\bf D56}, 5008, (1997).
\bibitem{sha1} 
B.J. Schaefer and H.J. Pirner, Nucl. Phys. {\bf A627}, 481, (1997); 
Nucl. Phys. {\bf A660} 439, (1999);
J. Meyer, G. Papp, H.J. Pirner and T. Kunihiro, Phys. Rev. 
{\bf C61}, 035202, (2000);
G. Papp, B.J. Schaefer, H.J. Pirner and J. Wambach, 
Phys. Rev. {\bf D61}, 096002, (2000);
O. Bohr, B.J. Schaefer and J. Wambach,
Int. J. Mod. Phys. {\bf A16}, 3823, (2001); 
Int. J. Mod. Phys. {\bf A16}, 2119, (2001); 
B. J. Schaefer, O. Bohr and  J. Wambach, {\it Finite temperature gluon condensate 
with renormalization group flow equations},
Preprint  Dec 2001, e-Print Archive: hep-th/0112087;
J. Meyer, K. Schwenzer, H.-J. Pirner and A. Deandrea,
Phys. Lett.  {\bf B526}, 79, (2002).
\bibitem{boza} A. Bonanno and  D. Zappal\`a, Phys. Lett.  {\bf B504}, 181, (2001).
\bibitem{maza} M. Mazza and  D. Zappal\`a, Phys. Rev.  {\bf D 64}, 105013, (2001). 
\bibitem{za} D. Zappal\`a, Phys. Lett.  {\bf A290}, 35, (2001).
\bibitem{liti1} D. F. Litim, Phys. Rev. {\bf D64}, 105007, (2001);  JHEP, 0111, 059, (2001);
\bibitem{liti2}D. F. Litim and J. M. Pawlowski, Phys. Lett.  {\bf B516}, 197, (2001). 
\bibitem{lpmain} D. F. Litim and J. M. Pawlowski, 
{\it Perturbation theory and renormalization group equations}
Preprint: CERN-TH-2001-327, FAU-TP3-01-10, Nov 2001, e-Print Archive: hep-th/0111191.
\bibitem{schw} J. Schwinger, Phys. Rev. {\bf 82}, 664, (1951).
\bibitem{pw}  T. Papenbrock and  C. Wetterich, Z. Phys. {\bf C65}, 519, (1995);
T. R. Morris, Nucl. Phys. {\bf B458}, 477, (1996);
M. Bonini, G. Marchesini and M. Simionato,  Nucl. Phys.  {\bf B483}, 475, (1997);
P. Kopietz, Nucl. Phys.  {\bf B595}, 493, (2001);
T. R. Morris and  J. F. Tighe, Int. J. Mod. Phys. {\bf A16}, 2095, (2001).
\bibitem{io2l} A. Bonanno and D. Zappal\`a, Phys. Rev. {\bf D57}, 7383 (1998);
A. Bonanno, V. Branchina, H. Mohrbach and D. Zappal\`a, Phys. Rev. {\bf D60}, 065009, (1999).
\end{thebibliography}
\enddocument